\documentclass{article}

\usepackage{PRIMEarxiv}

\usepackage[utf8]{inputenc} 
\usepackage[T1]{fontenc}    
\usepackage{hyperref}       
\usepackage{url}            
\usepackage{booktabs}       
\usepackage{amsfonts}       
\usepackage{nicefrac}       
\usepackage{microtype}      
\usepackage{lipsum}
\usepackage{fancyhdr}       
\usepackage{graphicx}       
\usepackage{multicol}
\usepackage{multirow}
\usepackage{array}
\usepackage{algorithm}
\usepackage{algorithmicx}
\usepackage{authblk}
\title{MDT-Net: Multi-domain Transfer by Perceptual Supervision for Unpaired Images in OCT Scan
}

\author[1,2]{Weinan Song}
\author[1]{Gaurav Fotedar}
\author[1]{Nima Tajbakhsh}
\author[1,2]{Ziheng Zhou}
\author[2]{Lei He}
\author[1,3]{Xiaowei Ding}

\affil[1]{VoxelCloud Inc}
\affil[2]{University of California, Los Angeles}
\affil[3]{Shanghai Jiao Tong University}

\begin{document}
\maketitle


\begin{abstract}
Deep learning models tend to underperform in the presence of domain shifts. Domain transfer has recently emerged as a promising approach wherein images exhibiting a domain shift are transformed into other domains for augmentation or adaptation. However, with the absence of paired and annotated images, models merely learned by adversarial loss and cycle consistency loss could result in poor consistency of anatomy structures during the translation. Additionally, the complexity of learning multi-domain transfer could significantly increase with the number of target domains and source images. In this paper, we propose a multi-domain transfer network, named \textit{MDT-Net}, to address the limitations above through perceptual supervision. Specifically, our model consists of a single encoder-decoder network and multiple domain-specific transfer modules to disentangle feature representations of the anatomy content and domain variance. Owing to this architecture, the model could significantly reduce the complexity when the translation is conducted among multiple domains. To demonstrate the performance of our method, we evaluate our model qualitatively and quantitatively on RETOUCH, an OCT dataset comprising scans from three different scanner devices (domains). Furthermore, we take the transfer results as additional training data for fluid segmentation to prove the advantage of our model indirectly, i.e., in the task of data adaptation and augmentation. Experimental results show that our method could bring universal improvement in these segmentation tasks, which demonstrates the effectiveness and efficiency of \textit{MDT-Net} in multi-domain transfer.


\end{abstract}

\section{Introduction}
Deep learning has proved effective in automating the diagnosis and quantification of various diseases and conditions. These models, however, tend to underperform in the presence of domain shifts, which commonly exist in medical imaging due to variance of scanner devices \cite{scannershift}, diversity of imaging protocols \cite{mri_adapt}, or deviation between real and synthesized data \cite{oral3d}. In the absence of paired and labeled data, models based on cycle consistent loss \cite{CycleGAN}\cite{segnorm} could provide a solution by learning from a round domain transfer process within two domains. However, such models are easy to lose the content consistency on diseased images (as shown in our experiment) during the transfer. Besides, the model can only learn a one-to-one domain transfer at one time, where the model complexity grows geometrically with the number of domains. In comparison, neural style transfer \cite{nst} provides a promising solution to keep the content consistency by aligning the statistical distribution of medical images collected from different sources \cite{nst_card}\cite{stylepath}\cite{nst_synthesis}, while the model complexity problem remains unsolved as the optimization time increases linearly with the number of images in source domains.

To address the limitations above, we introduce \textit{MDT-Net}, a multi-domain transfer model, to decouple the feature representation of anatomy structures and domain deviation of medical images with an encoder-decoder network and multiple domain transfer modules. Inspired by perceptual supervision \cite{perceptual}, \textit{MDT-Net} preserves anatomical structures during translation by imposing content loss in the identical domain transfer and domain loss (some may call style loss) in diverse domain transfer. Therefore, it can directly translate images into multiple target domains at one time without any reference images during the inference, which takes much less time than the methods based on neural style transfer. Moreover, the training complexity reduces from $\frac{n(n-1)}{2}$ to $n$ when involving transfer among $n$ domains compared with models based on CycleGAN. To demonstrate the translation performance, we first compare the translated results in variance against the target domains and similarity with the source content. Then we take these images as extra data to boost existing segmentation models for comparison of generation quality. Extensive results show that our model can significantly outperform other methods qualitatively and quantitatively.

\section{Methodologies}

\begin{figure}[t]
    \centering
    \includegraphics[width=\textwidth]{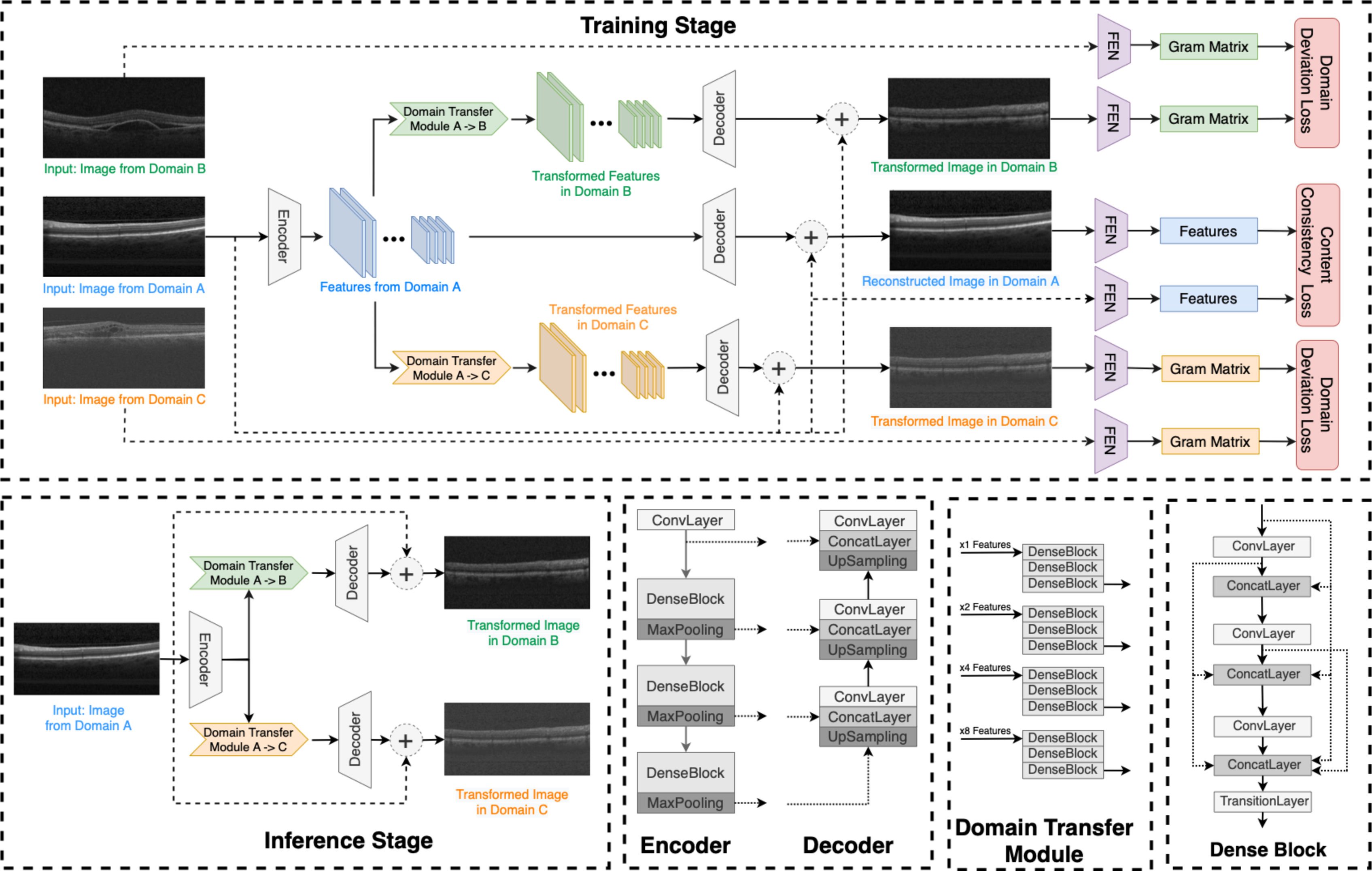}
    \caption{A figure illustration of our proposed model is shown in this picture. We use different colors to represent features and images in different domains, i.e., blue for the source domain, and orange and green for the two target domains.
    }
    \label{fig:model}
\end{figure}

\subsection{Keeping Anatomy Consistency during Translation}
As shown in \figurename~\ref{fig:model}, \textit{MDT-Net} consists of an encoder-decoder network ($f_e()$ and $f_d()$) to learn anatomy-consistent feature representation and multiple feature transfer modules ($t_i(), i=1,2\cdots,X$) to learn domain transition, where each module learns feature translation to a target domain. The training process during the translation is composed of two circumstances: 1) identical domain transfer, where the model generates an image $I^\prime$ by $f_d(f_e(I))$ from an image $I$ in the source domain,  and 2) diverse domain transfer, where the model outputs a translated image $I_X^\prime$ into the target domain $X$ via $f_d(t_X(f_e(I)))$. Since the domain transfer toward each target domain is learned explicitly by a feature transfer module, \textit{MDT-Net} can directly translate images into multiple target domains without any reference images by forwarding deep features into different feature transfer modules respectively during the inference.

\subsection{Perceptual Supervision}
Perceptual supervision is first proposed in \cite{nst} and has been widely applied in style transfer between paintings and photograph by capturing implicit content features and texture statistics. Generally, the perceptual loss is calculated by a pre-trained feature extraction network (FEN) $\mathcal{F}$ to compute the reconstruction loss over content and style features. In our model, we define the loss function as a combination of $\mathcal{L}_{content}$ and $\mathcal{L}_{domain}$ following the perceptual loss as:

\begin{equation}
\mathcal{L}=\mathcal{L}_{content}(I, I^\prime)+ \sum_{X}^{}\alpha_X\cdot\mathcal{L}_{domain}^{X}(I_X, I_X^\prime),
\end{equation}

where $I_X$ is the image randomly sampled in a target domain $X$. The content loss $\mathcal{L}_{content}$ and domain loss $\mathcal{L}_{domain}$ are defined as:

\begin{equation}
\mathcal{L}_{content}=\frac{1}{N_c} \sum_{l}^{l_1^c,\cdots ,l^c_{N_c} }\left\|\mathcal{F}^{l}(I^\prime)-\mathcal{F}^{l}(I)\right\|^{2}
\end{equation}

\begin{equation}
\mathcal{L}_{domain}=\frac{1}{N_d} \sum_{l}^{l_1^d,\cdots ,l^d_{N_d} }\left\|\mathcal{G}(\mathcal{F}^{l}(I^\prime_X))-\mathcal{G}(\mathcal{F}^{l}(I_X))\right\|^{2}.
\end{equation}

$\mathcal{F}^{l}(\cdot)$ denotes the features selected from the FEN and $\mathcal{G}(\cdot)$ is a function to compute the Gram matrix \cite{nst}, which has been widely used to compare the texture statistics in paintings.

\subsection{Network Architecture}
Our network architecture is developed based on StyleBank \cite{stylebank}. We make several improvements to accommodate it to the domain transfer problem in medical images: 1) We apply transfer modules on multi-level features generated by the encoding network to learn feature translation. 2) The transfer modules of \textit{MDT-Net} consist of multiple dense-connected convolution layers instead of a single convolution layer. 3) The model is trained to predict a residual image instead of the transfer result directly. Evaluation of these changes and generation comparisons between StyleBank and MDT-Net can be seen in the ablation study in section \ref{sec:results}. 

\section{Experiment}

\subsection{Dataset}
We use RETOUCH \cite{retouch} to validate the domain transfer capability of \textit{MDT-Net}. The dataset contains 70 OCT scans taken by three different vendors (domains): 1) 24 from Cirrus, 2) 24 from Spectralis, and 3) 22 from Topcon. For brevity, we use $C, S, T$ to represent each domain name. The images are annotated with three kinds of pixel-wise pathological annotations, which are only used for segmentation tasks in this paper.

\subsection{Evaluation}
We use Fréchet Inception Distance (FID) \cite{fid}\cite{fid_git} and Learned Perceptual Image Patch Similarity (LPIPS) \cite{lpips} to compare the domain similarity and content inconsistency of transferred images. We also propose Domain Perceptual Distance (DPD), a combination of FID and LPIPS, as an overall evaluation metric to indicate the distance to the optimal results, represented as:
\begin{equation}
{DPD}={FID} + \lambda\cdot{(1 - LPIPS)}\times 100\%.
\end{equation}
We use $\lambda=1$ in this paper. For segmentation, we use averaged dice scores of the three types of labels for comparison.

\subsection{Training}
We randomly select 5 cases from each domain in the dataset as test data and take the rest as training images in both domain transfer and segmentation experiments. We use VGG-16 \cite {vgg} as FEN and extract features from $relu4\_1$ for $\mathcal{L}_{C}$ and $relu1\_2, relu2\_2, relu3\_2, relu4\_2$ for $\mathcal{L}_{D}^{X}$. As \textit{MDT-Net} is a 1-to-n model, we need to train three models to generate all the transfer results. Due to the various number of B-scan slices in three domains, we train the model for 32, 80, and 40 epochs for $C$, $S$, and $T$ to balance training iterations. The training starts with a learning rate at $10^{-3}$ by Adam \cite{adam} and is decayed by $0.1$ in the middle.

\subsection{Comparison Models}
We compare \textit{MDT-Net} with four other unsupervised domain transfer models that are either based on cycle consistency, i.e., CycleGAN \cite{CycleGAN} and StarGAN\_v2 \cite{stargan2}, or neural style transfer, i.e., AdaIN \cite{adain} and StyleBank \cite{stylebank}. To cover all the six domain transfer circumstances, we train three models for CycleGAN (bidirectional 1-1), one model for StarGAN\_v2 (n-n), six models for AdaIN (taken as 1-1), and three models for Stylebank (1-n).

\section{Results}
\label{sec:results}

\begin{figure}[tp]
    \centering
    \includegraphics[width=\textwidth]{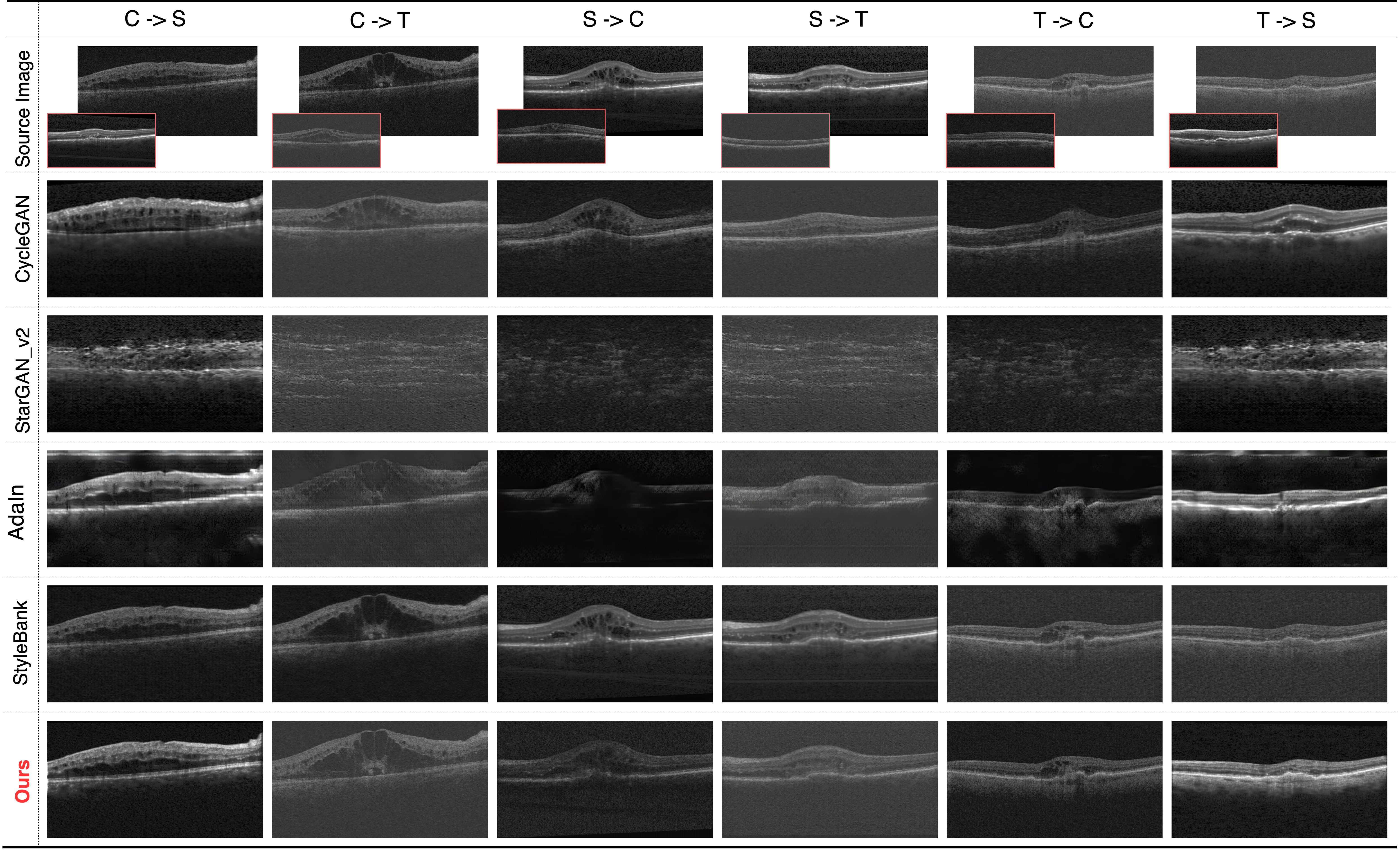}
    \caption{Qualitative comparison of transferred images generated by different algorithms in six types of domain transfer. Example images in the target domain (not used as reference) are shown with a red border in the bottom-left of each source image. \textit{MDT-Net} successfully transfers the images into various domains and perfectly preserves the anatomical structures, especially over AdaIN. As a comparison, CycleGAN has changed the retina structure, and StyleBank fails to transfer the images to the target domain. Additionally, StartGAN\_v2 hardly keeps the original content and can only generate textures that match the target domains.
    }
    \label{fig:transfer_results}
\end{figure}

\begin{figure}[tp]
    \centering
    \includegraphics[width=\textwidth]{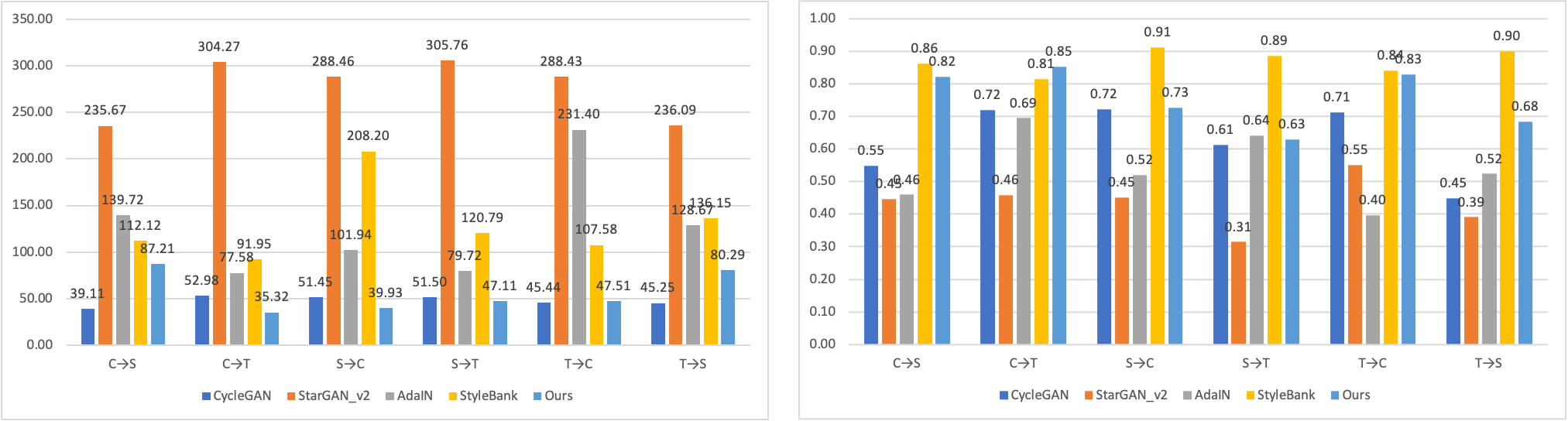}
    \caption{FID$\downarrow$ (left) and LPIPS$\uparrow$ (right) of transferred results in 6 types of transfer.}
    \label{fig:score}
\end{figure}

\subsection{Multi-domain Transfer}
We first compare the results generated by these five models in the task of domain transfer among three domains. We show qualitative results in \figurename ~\ref{fig:transfer_results} and averaged scores of FID, LPIPS, and DPD in Table~\ref{tab:transfer_compare}. Details of FID and LPIPS for each kind of domain transfer can be seen in \figurename ~\ref{fig:score}. We can see that our proposed method can achieve the best balance between domain shift and content consistency. 
Compared with \textit{MDT-Net}, CycleGAN can get excellent performance on domain shift, where the model fails to keep the anatomical structure of the retina. StyleBank preserves the content during transformation but can not reasonably match textures of target domains. As StarGAN\_v2 almost totally fails to reconstruct content, we exclude it in the latter experiments for fluid segmentation in terms of domain adaptation and data augmentation. To be noted, our model successfully achieves the domain transfer without losing details of the original content among all five models.

\begin{table}[tp]
    \centering
    \renewcommand{\arraystretch}{1.2}
    \caption{Averaged scores for six types of domain transfer by \textit{MDT-Net} and other four baseline models. \textit{MDT-Net} achieves the lowest DPD with the best trade-off between content consistency and domain transition.}
    \label{tab:transfer_compare}
    \setlength\tabcolsep{2pt}
    \begin{tabular}{p{1.8cm}<{\centering}p{1.8cm}<{\centering}p{2.2cm}<{\centering}p{1.8cm}<{\centering}p{1.8cm}<{\centering}p{1.6cm}<{\centering}}
    \hline
    Method&CycleGAN&StargGANv2&AdaIN&StyleBank&\textbf{Ours}\cr
    \hline
    FID$\downarrow$&47.62&276.45&126.51&129.47&56.23\cr
    LPIPS$\uparrow$ (\%)&62.72&43.48&53.88&86.85&75.67\cr
    \hline
    DPD$\downarrow$&84.91&332.96&172.62&142.62&\textbf{80.56}\cr
    \hline
    \end{tabular}
\end{table}

\subsection{Domain Adaptation}
This experiment assumes that only images in one domain are provided with pixel-wise annotations while the segmentation model is expected to adapt to the test data in the other two domains. This is a common issue in hospitals because labeled data and clinical images may come from different sources. We use DeeplabV3+ \cite{deeplabv3} as the baseline model for fluid segmentation. We first train the model with images from one domain, then add transferred images that share the same annotations as extra data. Therefore, the improved dice score brought by additional images can indicate the quality of transferred results. For example, for domain C, the difference of segmentation accuracy between the model trained with C only and the model trained with C, C $\rightarrow$ S and C $\rightarrow$ T is used to indicate the adaptation in unseen domains. As shown in Table~\ref{tab:exp_adapt}, \textit{MDT-Net} brings the biggest improvement and indicates the best transfer performance.

\begin{table}[tp]
    \centering
    \caption{Averaged segmentation results in the task of domain adaptation. For each domain, the segmentation model is trained with transferred results (except for baselines) in other domains. \textit{MDT-Net} brings the biggest improvement against other models.}
    \renewcommand{\arraystretch}{1.2}
    \label{tab:exp_adapt}
    \setlength\tabcolsep{1pt}
    \begin{tabular}{p{1.6cm}<{\centering}p{0.8cm}<{\centering}p{0.8cm}<{\centering}p{0.1cm}p{0.8cm}<{\centering}p{0.8cm}<{\centering}p{0.1cm}p{0.8cm}<{\centering}p{0.8cm}<{\centering}p{0.1cm}p{0.8cm}<{\centering}}
    \hline
    \multirow{2}{*}{Method}&\multicolumn{2}{c}{C}&&\multicolumn{2}{c}{S}&&\multicolumn{2}{c}{T}&&\multirow{2}{*}{Avg}\cr
    \cline{2-3}\cline{5-6}\cline{8-9}
    &S&T&&C&T&&C&S\cr
    \hline
    -&59.5&70.0&&59.1&64.8&&54.8&72.1&&63.4\cr
    CycleGAN&53.1&55.9&&52.2&60.8&&58.2&71.8&&58.7\cr
    AdaIN&67.2&77.6&&56.4&65.9&&77.5&77.3&&70.3\cr
    StyleBank&76.0&83.4&&62.5&84.4&&74.2&86.4&&77.8\cr
    \textbf{Ours}&84.5&76.9&&67.5&83.0&&77.3&85.5&&\textbf{79.1}\cr
    \hline
    \end{tabular}
\end{table}

\subsection{Data Augmentation}
In this experiment, we demonstrate the effectiveness of \textit{MDT-Net} by taking it as a data augmentation method. Different from the adaptation task where images from other domains and their transferred results are not seen during training, we use all images, including original images from three domains and their transferred results (3 + 6 types in all), as the training data for fluid segmentation. Similarly, we use improvements in segmentation accuracy to indicate the domain transfer performance. To avoid potential influence brought by variation in models, we introduce U-Net \cite{unet} and HR-Net \cite{hrnet} as additional baseline segmentation models. As shown in Table \ref{tab:exp_aug}, our method can best boost all three existing segmentation models, bringing about 2\% increase in dice scores.

\begin{table}[tp]
    \renewcommand{\arraystretch}{1.2}
    \centering
    \caption{Comparison of domain transfer models in the task of data augmentation. Our method best improves all three segmentation baselines, with about 2\% in dice scores.}
    \label{tab:exp_aug}
    \setlength\tabcolsep{1pt}
    \begin{tabular}{p{3.4cm}<{\centering}p{1.6cm}<{\centering}p{1.1cm}<{\centering}p{1.4cm}<{\centering}p{1.2cm}<{\centering}p{1.2cm}<{\centering}}
    \hline
    Segmentation Model&Method&C&S&T&Average\cr
    \hline
    \multirow{6}{*}{U-Net}
    &-&76.64&78.36&67.02&74.01\cr
    &CycleGAN&68.24&74.71&77.64&73.53\cr
    &AdaIn&69.55&73.78&78.18&73.84\cr
    &StyleBank&72.05&74.40&79.77&75.41\cr
    &\textbf{Ours}&83.97&71.74&73.35&\textbf{76.35}\cr

    \hline
    \multirow{6}{*}{Deeplab}
    &-&83.37&78.91&87.85&83.38\cr
    &CycleGAN&79.73&88.90&85.53&84.72\cr
    &AdaIn&80.66&87.59&85.86&84.70\cr
    &StyleBank&83.44&88.83&80.52&84.13\cr
    &\textbf{Ours}&81.61&89.16&87.63&\textbf{86.13}\cr

    \hline
    \multirow{6}{*}{HR-Net}
    &-&80.07&87.26&87.53&84.95\cr
    &CycleGAN&82.08&87.56&84.89&84.84\cr
    &AdaIn&82.20&80.58&87.97&83.58\cr
    &StyleBank&81.92&87.36&81.26&83.51\cr
    &\textbf{Ours}&81.46&89.87&87.42&\textbf{86.25}\cr
    \hline
    \end{tabular}
\end{table}

\subsection{Ablation Study}
\begin{figure}[tp]
    \centering
    \includegraphics[width=\textwidth]{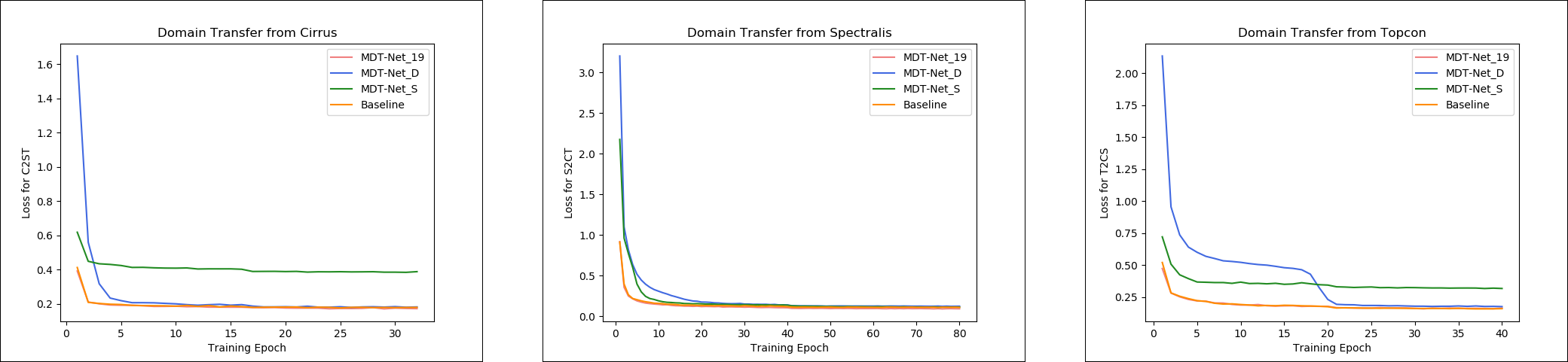}
    \caption{Change of the training loss of \textit{MDT-Net} in the ablation study.}
    \label{fig:ablation}
\end{figure}

In this experiment, we change the architecture of \textit{MDT-Net} for ablation study. \textit{MDT-Net\_D} removes the residual learning at the output, where the decoder directly generates the transferred results. \textit{MDT-Net\_S} replace the feature transformation module with the one proposed in StyleBank. \textit{MDT-Net\_19} takes VGG-19 as FEN. From Table. \ref{tab:exp_ablation} we can find that the proposed domain transfer module plays the most important role in our model. Combined with \figurename ~\ref{fig:ablation}, we can see that changing FEN does not affect the result while removing the residual structure mainly increases the converging time.

\begin{table}[tp]
    \centering
    \renewcommand{\arraystretch}{1.2}
    \caption{Averaged scores of transferred results from \textit{MDT-Net} in the ablation study.}
    \label{tab:exp_ablation}
    \setlength\tabcolsep{2pt}
    \begin{tabular}{p{1.8cm}<{\centering}p{2.2cm}<{\centering}p{2.2cm}<{\centering}p{2.2cm}<{\centering}p{2.2cm}<{\centering}}
    \hline
    Methods&\textit{MDT-Net\_D}&\textit{MDT-Net\_S}&\textit{MDT-Net\_19}&Baseline\cr
    \hline
    FID$\downarrow$&60.95&125.46&57.45&56.23\cr
    LPIPS$\uparrow$ (\%)&76.05&67.63&74.98&75.67\cr
    \hline
    DPD$\downarrow$&84.90&157.83&82.47&\textbf{80.56}\cr
    \hline
    \end{tabular}
\end{table}

\section{Conclusion}
In this paper, we introduce \textit{MDT-Net} to achieve multi-domain transfer within one single model trained by unpaired and unlabeled images with perceptual supervision. We disentangle the anatomy content and domain variance by an encoder-decoder network and multiple domain-specific transfer modules. Furthermore, extensive experiments on the task of transfer among three domains of OCT images have validated the advantage of \textit{MDT-Net} qualitatively and quantitatively.

\bibliographystyle{unsrt}  
\bibliography{ref}

\end{document}